\documentclass[english]{article}
\usepackage[T1]{fontenc}
\usepackage{color}
\usepackage{float}
\usepackage{graphicx}
\usepackage{esint}

\makeatletter
\newcommand{\lyxaddress}[1]{
	\par {\raggedright #1
	\vspace{1.4em}
	\noindent\par}
}

\usepackage{babel}
\usepackage[superscript,biblabel]{cite}

\makeatother

\usepackage{babel}
\begin{document}
\title{On the microscopic structure of neat linear alkylamine liquids: an
x-ray scattering and computer simulation study }
\author{Martina Po\v{z}ar$^{1}$, Lena Friedrich$^{2}$, Tristan Millet$^{3}$
\\
 Michael Paulus$^{2}$, Christian Sternemann$^{2}$\thanks{christian.sternemann@tu-dortmund.de}
and Aur\'{e}lien Perera$^{3}$ \thanks{aup@lptmc.jussieu.fr}}
\maketitle

\lyxaddress{$^{1}$Faculty of Science, University of Split, Ru{\dj}era Bo\v{s}kovi\'{c}a
33, 21000 Split, Croatia}

\lyxaddress{$^{2}$Fakult\"{a}t Physik/DELTA, Technische Universit\"{a}t Dortmund, D-44221
Dortmund, Germany}

\lyxaddress{$^{3}$Laboratoire de Physique Th\'{e}orique de la Mati\`{e}re Condens\'{e}e
(UMR CNRS 7600), Sorbonne Universit\'{e}, 4 Place Jussieu, F75252, Paris
cedex 05, France.}
\begin{abstract}
Ambient condition linear amines, from propylamine up to nonylamine,
are studied by x-ray scattering and Molecular Dynamics simulations
of various force field models. The major finding is that the pre-peak
in alkylamines is of about one order of magnitude weaker than that
in alkanols, hence suggesting much weaker hydrogen bonding induced
clustering of the amine groups than for the hydroxyl groups. Computer
simulation studies reveal that OPLS-UA model reproduces the pre-peak,
but with larger amplitudes, while the GROMOS-UA and CHARMM-AA force
fields show almost no pre-peak. Simulations of all models reveal the
existence of hydrogen bonded clusters, equally confirmed through the
prominent pre-peak of the structure factor between the nitrogen atoms.
But, this pre-peak gets nearly cancelled by the various combinations
of the atom-atom structure factors contributions to the scattering
intensity, except for the OPLS model. The purpose of this work is
to understand the weakness of the scattering pre-peak from the pair
correlation function perspective, considered as an order parameter
associated to the concept of charge order. The difference between
models is equally analyzed from the same perspective. The analysis
reveals the strong charge order induced structural similarity between
amines and water, as opposed to mono-ols. This is traced back to the
C2v symmetry of both the water molecule and the amine head group.
It explains both the existence of H-bonded clusters and the weak scattering
pre-peak. Concerning the models, the presence or absence of partial
charges in the methyl groups of the alkyl tails explains the presence
or absence of the pre-peak in the calculated scattering intensities 
\end{abstract}

\section{Introduction}

Several recent studies of alcohols have put forward the rich variety
of hydrogen bonded hydroxyl group aggregates, both from the x-ray
scattering experiments \cite{EXP_Scatt_Magini_Methanol,EXP_Scatt_Narten_EthMeth,EXP_Sarkar_Joarder_Methanol,EXP_Sarkar_Joarder_ethanol,EXP_Scatt_Finns_Monools,EXP_Joarder_Alcohols,EXP_SIM_Wat1Prop,EXP_Scatt_Matija_Monools,EXP_SIM_Tomsic_butanol}
and computer simulations \cite{SIM_Bako_Neat_Meth_Clusters,SIM_Bako_methanol_EXP,SIM_Ludwig_Methanol,EXP_SIM_BenmoreEthanol,SIM_Finci2_linear_alcohols,2020_Linear_alcohols,2021_Octanols}.
The latter studies have demonstrated the chain-like association patterns
of the OH groups, witnessing plain and branched chains, as well as
loops and lassos \cite{2020_Linear_alcohols,2021_Octanols}. These
studies have also highlighted the importance of the alkyl chains,
which are not simply low energy inert components which follow the
association tendencies of the high energy polar head groups, but they
contribute entropically to reduced and condition the association of
the hydroxyl heads. Since self-assembly and self-association are not
only limited to hydrogen bonding between the OH groups, it is interesting
to investigate whether other forms of associations are equally rich
in patterns. The next in the list is the amine group NH2, which is
somewhat reminiscent of the water OH2 geometry. Yet, it also reminds
that, whereas water is the ``mother'' of the hydrogen bonding liquids
\cite{Water_Science_Frank1970,Water_Science_Stillinger1980,Water_Hbond_Ohmine1993,Water_Nature_Stanley_Mishima1998,Water_EXP_HeadGordon2002,Water_Stanley2008,Water_review_Gallo2016},
interestingly enough, it lacks the demonstrative hydrogen bonding
patterns that the alcohols show. Indeed, whereas alcohols demonstrate
the existence of aggregates in the x-ray scattering through the pre-peak
feature \cite{EXP_Scatt_Narten_EthMeth,EXP_Scatt_Finns_Monools,EXP_Scatt_Matija_Monools,2020_Linear_alcohols},
water has no such pre-peak, but a weak shoulder pattern \cite{Water_Perera2011}.
This lack of aggregate signature is equally observed in cluster distributions
calculated from computer simulations: all alcohols show an universal
OH pentamer leading cluster pattern, while water has quite a featureless
cluster distribution not very different from that of a standard Lennard-Jones
liquid.

Interestingly, unlike alkanols, smaller alkylamines such as methylamine
and ethylamines are not liquid in ambient conditions. This is already
an indication that the hydrogen bonding between the amines groups
alone is not able to stabilize a dense liquid, unlike the hydroxyl
groups for very small alkyl chains. An indirect conclusion is that
longer alkyl chains help stabilize the liquid state starting from
proplylamine. This conclusion was not obvious from the previous studies
of alkanols. Indeed, most short alkanes, such as methane to butane,
are also not liquids in ambient conditions \cite{Textbook_GorzynskiSmith2010}.
All this points to the essential role of the hydrophobic tails when
combined with hydrogen bonding head groups.

Another interesting issue is that of the model representation of the
molecular liquids when dealing with computer simulations. Several
studies have highlighted the importance of studying various force
field models \cite{2020_Linear_alcohols,2021_Octanols}. For instance
the scattering pre-peak shape for linear monools differs appreciably
from one force model to another. This is quite simply related to small
differences in the geometrical packings of the various charged groups,
which we call charge ordering, which add up in the structure correlation
functions and reflect upon the total scattering intensity, since the
latter is simply a sum over all partial atom-atom structure factors.

Following our recent work on x-ray scattering and simulation study
of alkanols \cite{2020_Linear_alcohols}, we study amines, which show
quite weak x-ray scattering pre-peak. We are concerned with the microscopic
reason for this property, particularly as revealed through pair correlation
functions and associated structure factors. These two quantities although
not directly available through experiments, participate to the x-ray
scattering, hence help explain its feature. However, they are accessible
through simulation and are biased by the choice of force field models.
We try to obtain explanations that are independent of this bias.

The literature is scarce when it comes to either experimental or simulations
studies of the structure of liquid amines. The earliest work on x-ray
diffraction appears to be that of Thosar \cite{old-amine-JCP} in
1938, who reports invariance of the main peaks positions for various
amines. This feature is now trivially explained in terms of the van
der Waals radii similarities between the carbon and nitrogen atoms.
Excluding the previous work of some of the authors \cite{2019_Propylamine2},
there are equally X-ray scattering experimental results on solid state
amines \cite{Amines_Solid_Xray_Maloney2014,Amines_Solid_Sacharczuk2023}.
As for the simulations, studies that feature the structuring in neat
amines include that of Kusalik and coworkers \cite{Amines_SIM_kusalik2000},
Kosztolányi et al. \cite{Amines_SIM_Bako_Palinkas2003} and Bauer
and Patel \cite{Amines_SIM_Charmm_Bauer}. The former two papers describe
the structure in methylamine, while the latter paper of Bauer and
Patel \cite{Amines_SIM_Charmm_Bauer} features methylamine, ethylamine
and propylamine. The group of Lachet did extensive work on force field
development of primary, secondary and tertiary amines \cite{Amines_Lachet_FF,Amines_Lachet_FF_secondary_amines},
with subsequent investigations about the transport properties of amines
\cite{Amines_Lachet_transport_equilibrium} and gas solubility in
amines \cite{Amines_Lachet_SIM_gas_solubility}. There were also simulation
studies about transport properties \cite{Amines_SIM_transport_Jadran}
and diffusion of amines \cite{Amines_SIM_diffusion_Orozco} undertaken
by other authors. However, the remainder of the literature of amine
simulations has the overarching theme of industrial applications in
the context of $\mathrm{CO_{2}}$ capture \cite{Amines_CO2_narimani2017,Amines_CO2_sharif2020,Amines_CO2_sinehbaghizadeh2023}
and the development and improvement of various materials \cite{Amines_SIM_Materials2005,Amines_Resin_estridge2018,Amines_SIM_materials2024}.

\section{Technical considerations}

\subsection{Experimental setup}

The wide-angle x-ray diffraction experiment was performed at beamline
BL9 of the DELTA synchrotron radiation source (Dortmund, Germany)
\cite{Krywka_EXP}. Propylamine (99\%) butylamine (99.5\%), pentylamine
(99\%), hexylamine (99\%), heptylamine (99\%), octylamine ($\geq$
99.5\%) and nonylamine (MQ200) were purchased from Sigma-Aldrich and
used without further treatment. The liquids were filled into borosilicate
glass capillaries of 1.5 mm diameter prior to the measurements and
capillaries were sealed. The incident photon energy was set to 20
keV which refers to a wavelength of 0.61992$\mathring{A}$ and the
scattered intensity was measured using a MAR345 image plate scanner.
The setup was calibrated with the diffraction image of a CeO$_{2}$
reference sample. In order to better assess the air scattering background
because of the weak pre-peak signals, diffraction data were taken
with and without a He beam path placed behind the sample holder. The
measured diffraction images were integrated using the program package
Fit2D \cite{Hammersley_EXP}. Finally the background due to scattering
from air and the glass capillary was subtracted and the diffraction
patterns were normalized by their integral in the momentum-transfer
$k$ range between 0.37 and 3 $\mathring{A}^{-1}$ to their calculated
counterparts, i.e. the averaged integral value of all calculated diffraction
intensities for a certain sample. Except for nonylamine, diffraction
data in the $k$-range between 0.035 and 0.3 $\mathring{A}{}^{-1}$
has been measured independently at the small angle x-ray scattering
beamline BL2 of DELTA using the setup as describe in Ref. \cite{Dargasz_2022}
with an incident energy of 12 keV, i.e. a wavelength of 1.0332$\mathring{A}$.
The calibration of the setup was performed with a silver behenate
reference and the raw data were treated as discussed for the wide-angle
x-ray scattering measurements. For representation, both data taken
at BL9 and BL2 were merged using a scaling factor in the $k$-range
were both sets overlap

\subsection{Computer simulation and theoretical details}

We performed molecular dynamics simulations of neat amines in the
Gromacs program package \cite{MD_Gromacs}. In our previous work \cite{2017_Propylamine1,2019_Propylamine2},
we chose the force field Gromos 53a6 \cite{FF_Gromos_53a6} for propylamine,
courtesy of the Automated Topology Builder \cite{FF_Gromos_ATB}.
Due to our previous experiences with linear alcohols \cite{2020_Linear_alcohols},
we examined different force fields: the CHARMM all-atom \cite{FF_Charmm1,FF_Charmm2,FF_Charmm3},
GROMOS 54a7 united-atom \cite{FF_Gromos_54a7} (from butylamine to
octylamine) and OPLS united-atom \cite{FF_OPLS_AA_organic_liquids_amines,FF_OPLS_alcohols_1}.

PACKMOL \cite{MD_Packmol} was used to create the initial configurations
of 2048 molecules for all amines, which underwent energy minimization
and equilibration for 4 ns. Production runs of 2 ns were performed,
during which at least 2000 configurations were collected. The simulations
were done in the NpT ensemble at ambient conditions, \textit{T} =
300 K and \textit{p} = 1 bar. The temperature was maintained with
the v-rescale thermostat \cite{MD_thermo_Vrescale}, whereas the the
Parrinello-Rahman barostat \cite{MD_barostat_Parrinello_Rahman_1,MD_barostat_Parrinello_Rahman_2}
was used to keep the pressure constant. The temperature algorithm
had a time constant of 0.2 ps and the pressure algorithm was set at
2 ps.

The integration algorithm was leap-frog \cite{MD_INT_Leapfrog}, which
had the time step of 2 fs. The short-range interactions were calculated
within the 1.5 nm cut-off radius, while the long-range electrostatics
were handled with the PME (Partial Mesh Ewald) method \cite{MD_PME}.
The LINCS algorithm \cite{MD_constraint_LINCS} handled the constraints.

The scattering intensity was calculated via the Debye formula \cite{Debye1,Debye2}:

\begin{equation}
I(k)=r_{0}^{2}\rho\sum_{ij}f_{i}(k)f_{j}(k)S_{ij}^{(T)}(k)\label{Ik}
\end{equation}
where $\rho=N/V$ is the density (where N is the number of molecules
in the volume V), the $f_{i}(k)$ functions are the form factor of
atom $i$, and $r_{0}=2.8179$ $\cdot10^{-13}$cm is the electronic
radius. The total structure factor $S_{ij}^{(T)}(k)$ is defined as
\cite{2020_Linear_alcohols}: 
\begin{equation}
S_{ij}^{(T)}(k)=w_{ij}(k)+\rho S_{ij}(k)\label{ST(k)}
\end{equation}
where $w_{ij}(k)$ the intra-molecular atom-atom structure factor,
$S_{ij}(k)$ is the atom-atom structure factor defined as the Fourier
transform of the atom-atom intermolecular pair correlation function
$g_{ij}(r)$ 
\begin{equation}
S_{ij}(k)=\rho\int d\vec{r}\left[g_{ij}(r)-1\right]\exp(i\vec{k}\cdot\vec{r})\label{Sk}
\end{equation}

\section{About the very small scattering pre-peak in relation to the amine
group clustering}

\subsection{X-ray scattering}

\begin{figure}[H]
\includegraphics[width=0.9\textwidth]{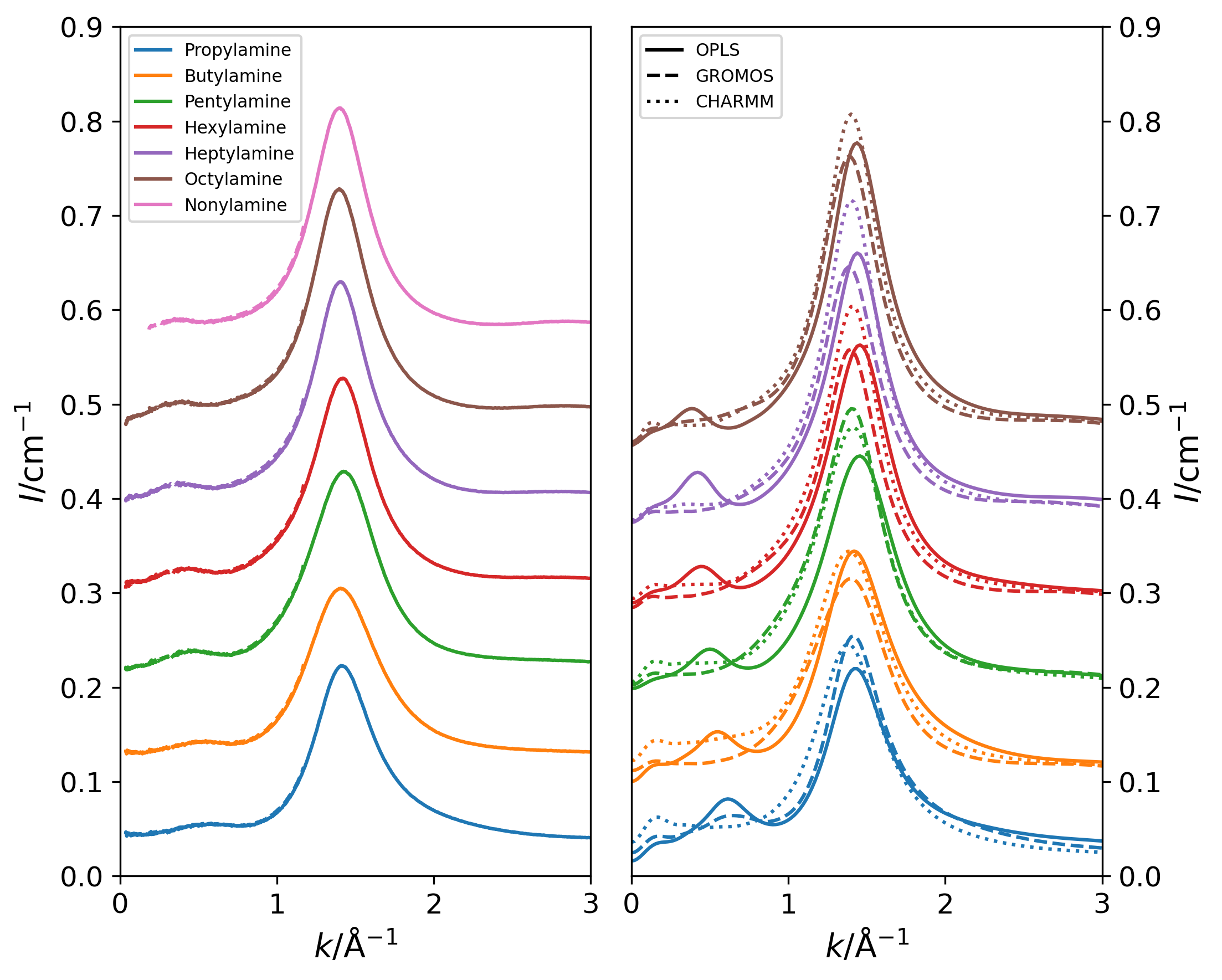}

\caption{X-ray scattering intensities $I(k)$ for all alkylamines (each shifted
by 0.1 cm$^{-1}$): from experimental data (left panel) and from computer
simulations of the three models (right panel). The dashed lines in
the experimental data show the results measured with He beam path
while small dots show the data measured in the small angle scattering
regime. The OPLS force field data is shown in full lines,
The CHARMM data in dotted lines and the GROMOS data in dashed lines.}

\label{Fig-Ik-comp} 
\end{figure}

Fig.\ref{Fig-Ik-comp} shows a comparison of the x-ray scattering
intensity $I(k)$ between the experimental data for all alkylamines
(left panel) and the $I(k)$ calculated from different simulation
models (right panel).

Very small experimental scattering pre-peaks are observed in contrast
with the relatively high ones observed for mono-ols\cite{2020_Linear_alcohols}.
Of the simulation data, only the OPLS model displays clear pre-peaks
with trends similar to that from experiments but with larger amplitude.

\begin{figure}[H]
\includegraphics[scale=0.35]{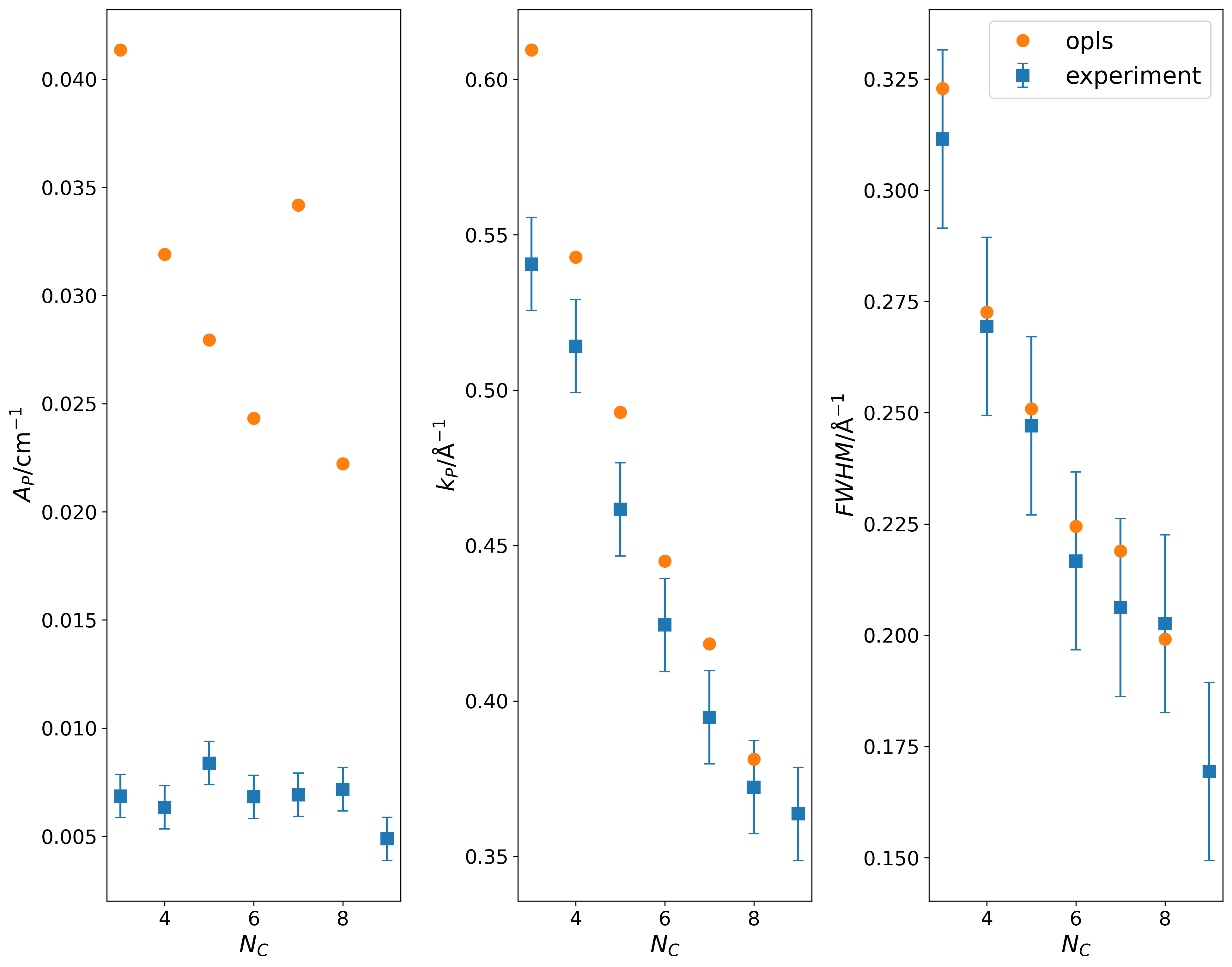}

\caption{Comparison of pre-peak characteristics with variation of chain length
$N_{C}$ between experimental and OPLS results for pre-peak. The left
panel shows the pre-peak amplitudes $A_{P}$, the central panel for
the positions $k_{P}$ and teh right panel for the full witdth at
half maximum. }

\label{Fig-PP-analysis} 
\end{figure}

In order to analyze the pre-peak characteristics, we subtracted the
main-peak tail and then fitted the pre-peak by a Pearson VII function
to extract its amplitudes $A_{P}$ , pre-peak position $k_{P}$ and
Full Width at Half Maximum (FWHM) applying the same procedures as
done in Ref.\cite{2020_Linear_alcohols} for proper comparison with
the mono-ols. The results of this analysis for the experimental data
and the results of the OPLS model are presented in Fig.\ref{Fig-PP-analysis}.
The error bars were determined based on the systematic error caused
by the background subtraction using both the measurements with the
He beam path and the ones in air.

The experimental data show clear trends of the x-ray spectra when
going from propylamine to nonylamine. The main peak intensity slightly
increases with increasing length of the carbon chain $N_{C}$ while
its position can be found around 1.41 $\mathring{A}$$^{-1}$ for
all amines. As evident from Fig.\ref{Fig-PP-analysis},the
position of the pre-peak shifts to smaller wave vector transfer $k_{P}$
from $0.54$ to $0.36$ $\mathring{A}$$^{-1}$ with $N_{C}$ . The
amplitude of the pre-peaks lies around $0.007$ cm$^{-1}$ while their
FWHM significantly decreases. These trends are similar to that observed
in mono-ols, except for the pre-peak amplitude difference of nearly
one order of magnitude, which we will discuss this in the next sections.
These trends can be explained similarly to that for alcohols\cite{2020_Linear_alcohols,2021_Octanols}.
It is the size of the methylene group that dominates the main peak,
hence explaining both the position and amplitude (proportional to
the number of carbons) using simple Bragg law argument. A similar
argument on the size of the hydrogen bonded aggregates explains the
pre-peak position, which tend to increase in size with longer amines,
while the concentration of the aggregates tends to weakly diminish,
considering the changes in the amplitudes and the FWHM. Particularly
for the smaller amines, the pre-peak positions of the corresponding
mono-ols are found to be at larger $k_{P}$ which can be assigned
to the larger polar groups in case of the amines.

Of the simulation data, only the OPLS model displays clear pre-peaks
with trends similar to that from experiments, and consistent with
that obtained with the same model in the case of the mono-ols. These
pre-peaks are smaller than those of the mono-ols, but still a factor
5 (butylamine) to 3 (octylamine) too high when compared with the experimental
data. The trend in pre-peak position with increasing chain length
is reasonably well reproduced by the OPLS model. The fact that the
simulation spectra are not in so much good agreement with the experimental
ones as it is for monols\cite{2020_Linear_alcohols} will be debated
in the Discussion section 7.

In contrast, both CHARMM and GROMOS force fields do not show clear
pre-peaks, except for the propylamine GROMOS data. The CHARMM
model being an all-atom model, one would expect the data to be closer
to that from experiments, but this does not appear to be case in a
clear fashion.

How to explain this near absence of pre-peaks in these last 2 models,
as well as the contrasting results across models? In fact, it is the
weakness of the pre-peak in the experiments that guide the interpretation.
Because the main question is rather: why the pre-peak for amines is
so small when compared to the mono-ols? Indeed, the weakness of the
pre-peak might be due to weak hydrogen bond clustering of the amine
head group when compared with that of the hydroxyl group. In this
case, some models might exaggerate this clustering while other might
totally underestimate it. The answer is neither, and lies on the nature
of the charge ordering, as we will explain in Section 4.

\subsection{Clustering of the amine groups}

The clustering of the amine groups can be studied from computer simulations,
either by direct probing of the cluster distribution probability $P_{n}(s)$
for cluster size $s$ of amine alkyl chain rank $n$, or by looking
at snapshots from various spatial configurations, and finally by analysing
the pair distribution functions of the amine atoms and the corresponding
structure factor.

\begin{figure}[H]
\includegraphics[width=0.8\textwidth]{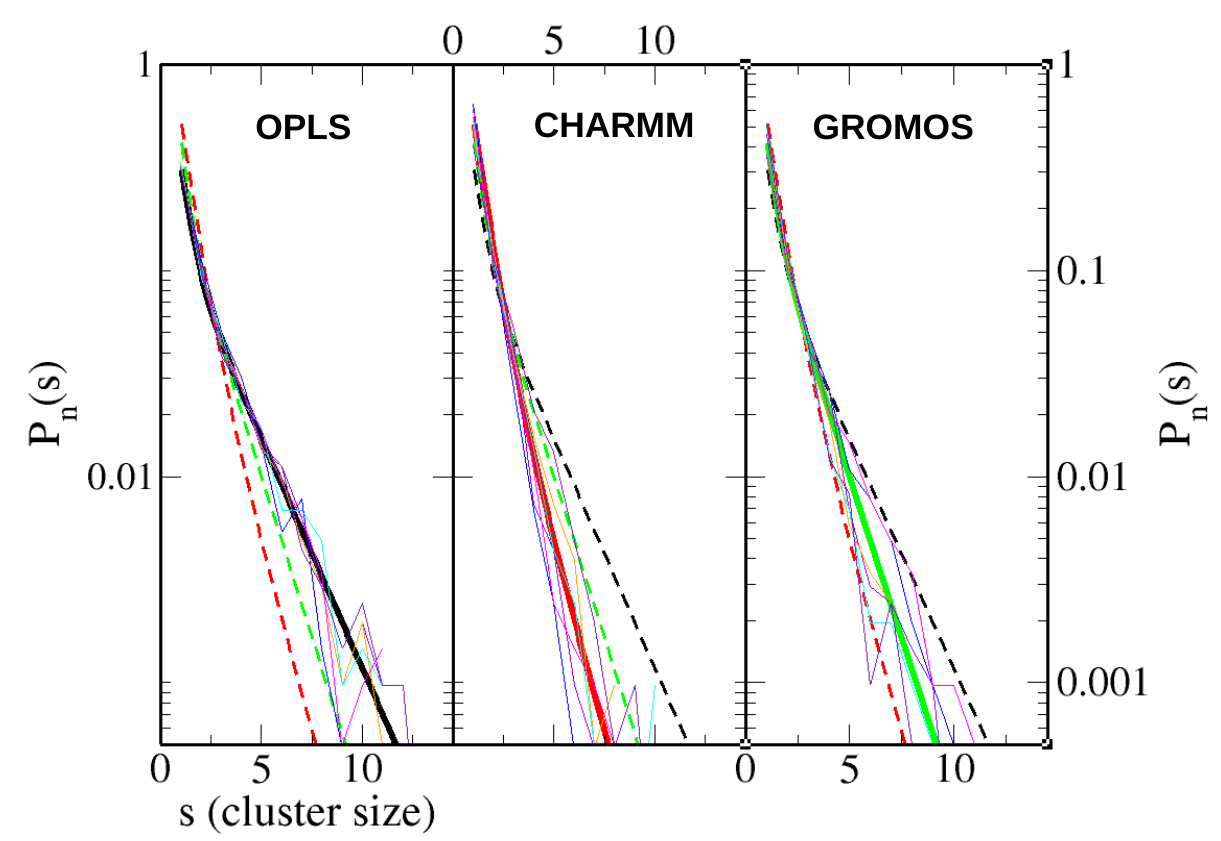}

\caption{Cluster size $s$ probability distribution functions $P_{n}(s)$ for
each model in separate panels, and for different alkylamines ($n=3,..,8$)
in each of the panels. The fits from Eq.(\ref{Pn}) are shown as thick
lines, black for OPLS, red for CHARMM and green for GROMOS. Full lines
are used for the model represented in the panel, with dashed curves
for the two other models.}

\label{FigClust} 
\end{figure}

Fig.\ref{FigClust} shows the $P_{n}(s)$ as function of the cluster
size $s$, for different alkylamines ($n=3,..,8$) and different force
field models. The important difference with mono-ols is that these
cluster probability distributions do not have the characteristic pentamer
maximum found for all mono-ols\cite{2020_Linear_alcohols,2021_Octanols}.
Instead, we find an exponential shape very similar to that of water\cite{AUP_Neat_Alcohols_JCP,AUP_Neat_alcohols_PRE}.
This finding could be related to the geometrical proximity of the
Y-shaped polar heads, $NH_{2}$ for amines and $OH_{2}$ for water\textcolor{blue}{,
}which importance will be discussed later in the paper.

A closer look at Fig.\ref{FigClust} shows that, despite the very
small differences between models, the OPLS model has a general tendency
to have less monomers and more higher n-mers than the two other models,
which can then be related to the trends of this model to enhance the
scattering pre-peak. Conversely, the CHARMM model is that which has
lowest overall cluster probabilities, while GROMOS is intermediate.
This hierarchy between the 3 models is further exemplified by the
2-exponential fits of Eq.\ref{Pn}, shown in Fig.\ref{FigClust} in
thick lines, the black line for OPLS being above the GROMOS green
curve and the CHARMM red curve.

The cluster distribution data can be fitted to a two exponential form:
\begin{equation}
P_{n}(s)=\exp(a_{0}+a_{1}s)+\exp(a_{2}+a_{3}s)\label{Pn}
\end{equation}
where the $a_{k}$ coefficients are given in Table 7 of the SI document.

\subsection{Typical cluster shapes}

From the computer simulations we can extract typical cluster shapes
as given by our cluster extraction program. Most abundant clusters
are dimers of the hydrogen bonded amine groups, such as that in Fig.\ref{FigClust}.
But there are also larger clusters, as shown in Fig.(\ref{FigClust}),
although their probability decrease exponentially (See Eq.(\ref{Pn})).
The figure shows how the alkyl tails decorrelate from the amine cluster
backbone.

\begin{figure}[H]
\includegraphics[width=0.9\textwidth]{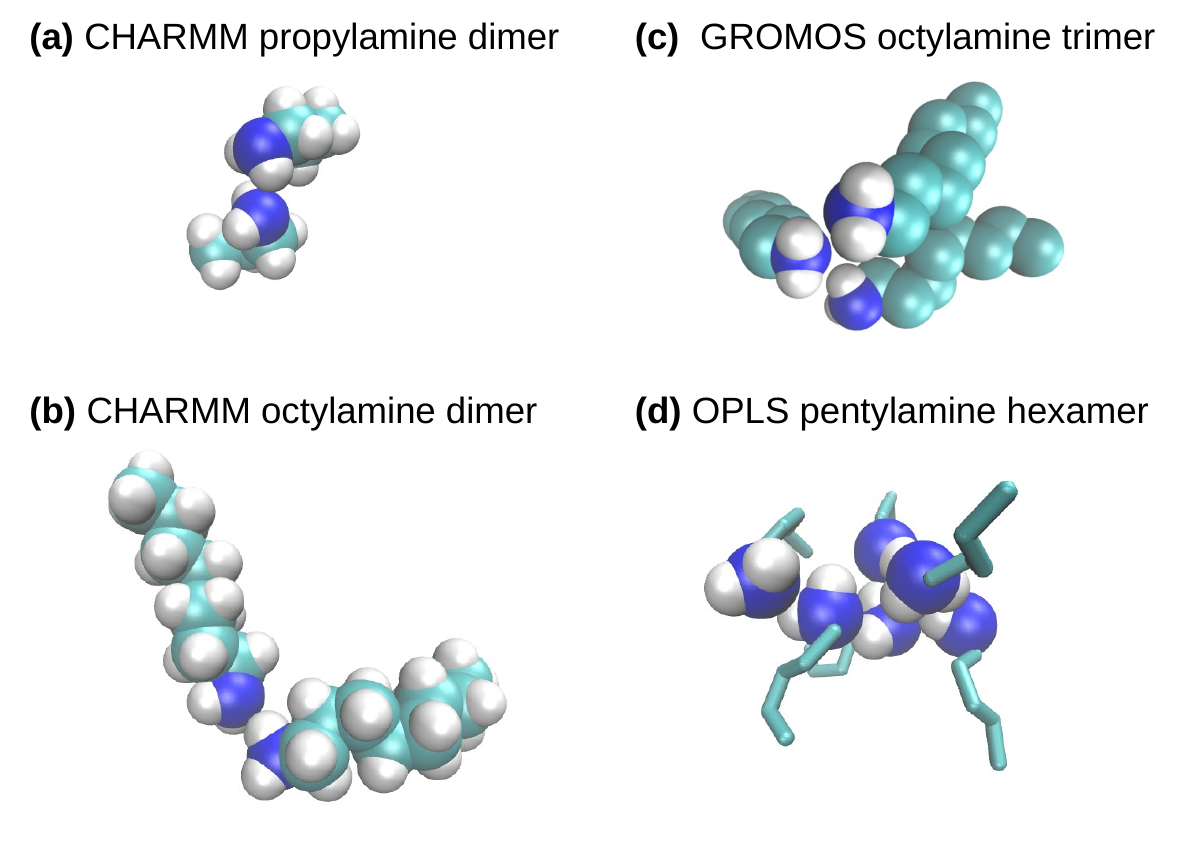}

\caption{Typical clusters for different force field models. In (d), the alkyl
tail is represented in the VMD licorice convention in order not to
block the view of the amine head group chain cluster. }
\end{figure}

It is also interesting to observe the difference between the bulkiness
of the tails in all-atom modeling, compared with that of united atom.
These figures also show how the hydrogen bonding with amines is not
similar to water, because of the presence of the alkyl group which
prevent tetrahedral distribution of the amine head groups. This is
also probably the reason why clusters are small. It is also seen that
chain clusters of the amines are not the only possibility, and that
bulky groups exists, such as that in (c). OPLS has more clear chain
clustering, as seen in (d).

\section{Charge order analysis through the order parameter functions}

In disordered liquids, the order parameter (in the sense of the Landau
free energy\cite{Landau:1980mil}) is the density\cite{Chaikin_Lubensky_1995}
, which is just a number $\rho_{a}=N_{a}/V$ ($N_{a}$ is the number
of particle of species $a$ per volume $V$), and will not provide
any information about the microscopic structure of the liquids. Yet,
associating liquids exhibit strong local order, and this shortcoming
from the theoretical perspective might seem disappointing at first.
However, the density $\rho$ is in fact the one-body function $\rho_{a}^{(1)}(\mathbf{x}$)
for species $a$, which has a meaning for interfacial or orientationally
ordered liquids since these type of systems require a general spatial
variable $\mathbf{x}$, the latter which represents the position and/or
orientation of a particle, but is absent in disordered liquids since
there is no preferred position or orientation. In the absence of a
meaningful one-body function, one may turn to the next function in
the hierarchy, which is the two-body $\rho_{ab}^{(2)}(\mathbf{x}_{1},\mathbf{x}_{2})$
for a pair of species $a$ and $b$, which formally defines the pair
correlation function $g_{ab}(\mathbf{x}_{1},\mathbf{x}_{2})$ by the
relation\cite{Textbook_Hansen_McDonald} 
\begin{equation}
\rho_{ab}^{(2)}(\mathbf{x}_{1},\mathbf{x}_{2})=\rho_{a}^{(1)}(\mathbf{x}_{1})\rho_{b}^{(1)}(\mathbf{x}_{2})g_{ab}(\mathbf{x}_{1},\mathbf{x}_{2})\label{2bodyGEN}
\end{equation}
but for the type of usual disordered liquids becomes 
\begin{equation}
\rho_{ab}^{(2)}(\mathbf{x}_{1},\mathbf{x}_{2})=\rho_{a}\rho_{b}g_{ab}(r)\label{2body}
\end{equation}
where $r=|\mathbf{r}_{1}-\mathbf{r}_{2}|$ is the relative distance
between the 2 particles (where we consider only distances $\mathbf{r}$
without including orientations as with $\mathbf{x}$ variables). In
case of disordered molecular liquids, the order parameter $g_{ab}(r)$
represents the pair distribution function between 2 atoms $a$ and
$b$, where it is assumed to consider the molecular liquid as a ``soup''
of atoms, and the intra-molecular part $w_{ab}(r)$ has been introduced
in Section 2.2. This way, the molecular liquid is seen ordered through
the charges born by the atoms, constrained by the intra-molecular
bonds.

The hydrogen bonding between the amine groups $NH_{2}$ are then simply
a Coulomb association between two negatively charged nitrogen atoms
$N$ through one of the positively charged hydrogen atom $H$. In
that, the process is similar to hydrogen bonding between the hydroxyl
$OH$ group in alkanols\cite{2020_Linear_alcohols}. Charge order
is then a classical version of the quantum physics of the hydrogen
bonding between 2 molecules. However, it also allows to described
in a unified way both the molecular hydrogen bonding and the Coulomb
association with ionic species. This approach is conceptually supported
by the fact that the order parameter function $g_{ab}(r)$ has a characteristic
shape for atoms that are associated through charge ordering, depending
of the nature of the valences\cite{PERERA2015243} .

\subsection{Charge order and scattering}

\begin{figure}[H]
\includegraphics[width=0.9\textwidth]{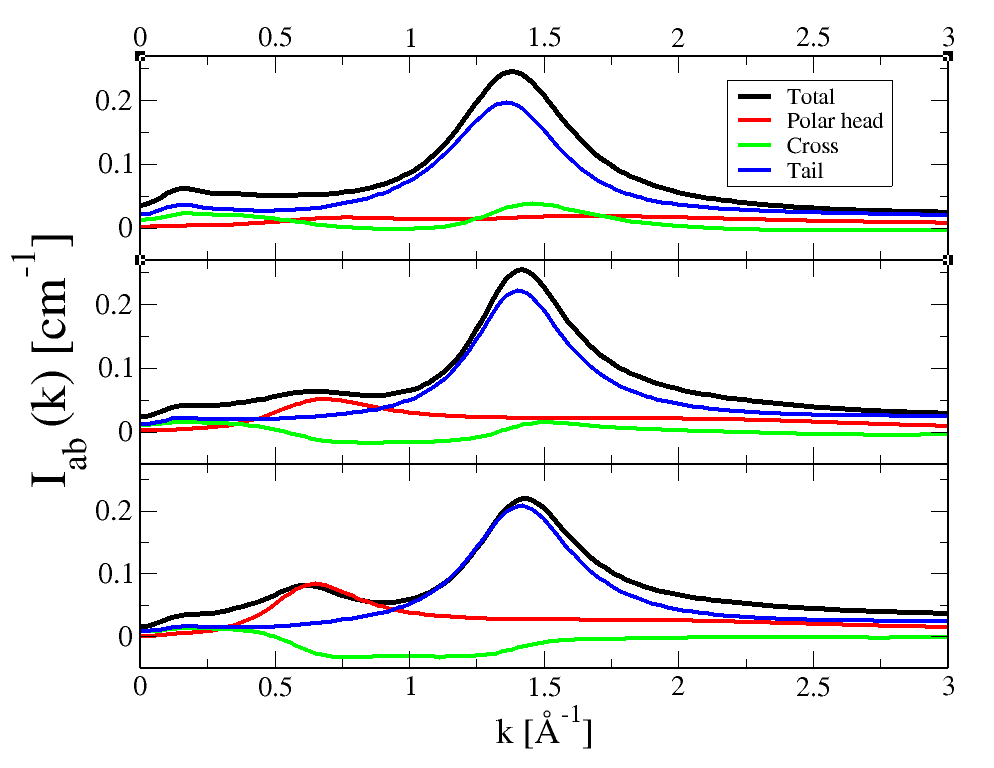}

\caption{Partial scattering contributions for propylamine }

\label{FigPartialScatt3} 
\end{figure}

In order to demonstrate the influence of charge ordering on x-ray
scattering intensity $I(k)$, we separate in Eq.(\ref{Ik}) the amine
polar head group contributions $I_{AA}(k)$ from that of the alkyl
tail $I_{TT}(k)$, as well as the cross contributions $I_{AT}(k)$,
with 
\begin{equation}
I(k)=I_{AA}(k)+I_{AT}(k)+I_{TT}(k)\label{Ik-sep}
\end{equation}
These 3 contributions are shown in Fig.(\ref{FigPartialScatt3}) for
propylamine, together with the total intensity $I(k)$(black curves).

Since propylamine is the only amine which shows a clear weak pre-peak
in x-ray scattering for all force field models (but this is not true
for CHARMM), it is interesting to see which part contributes mostly
to it. This is particularly visible for the GROMOS and the OPLS force
fields, where it is the polar head contribution which dominates, while
the cross contributions tend to be negative around $1\mathring{A}^{-1}$.
We also note that it is the alkyl tail which contributes essentially
to the main peak, since it contains most atoms, with size $\sigma_{CH_{2}}\approx4.5\mathring{A}$
corresponding to the main peak position $k=2\pi/\sigma_{CH_{2}}\approx1.4\mathring{A}^{-1}$. 

\begin{figure}[H]
\includegraphics[width=0.9\textwidth]{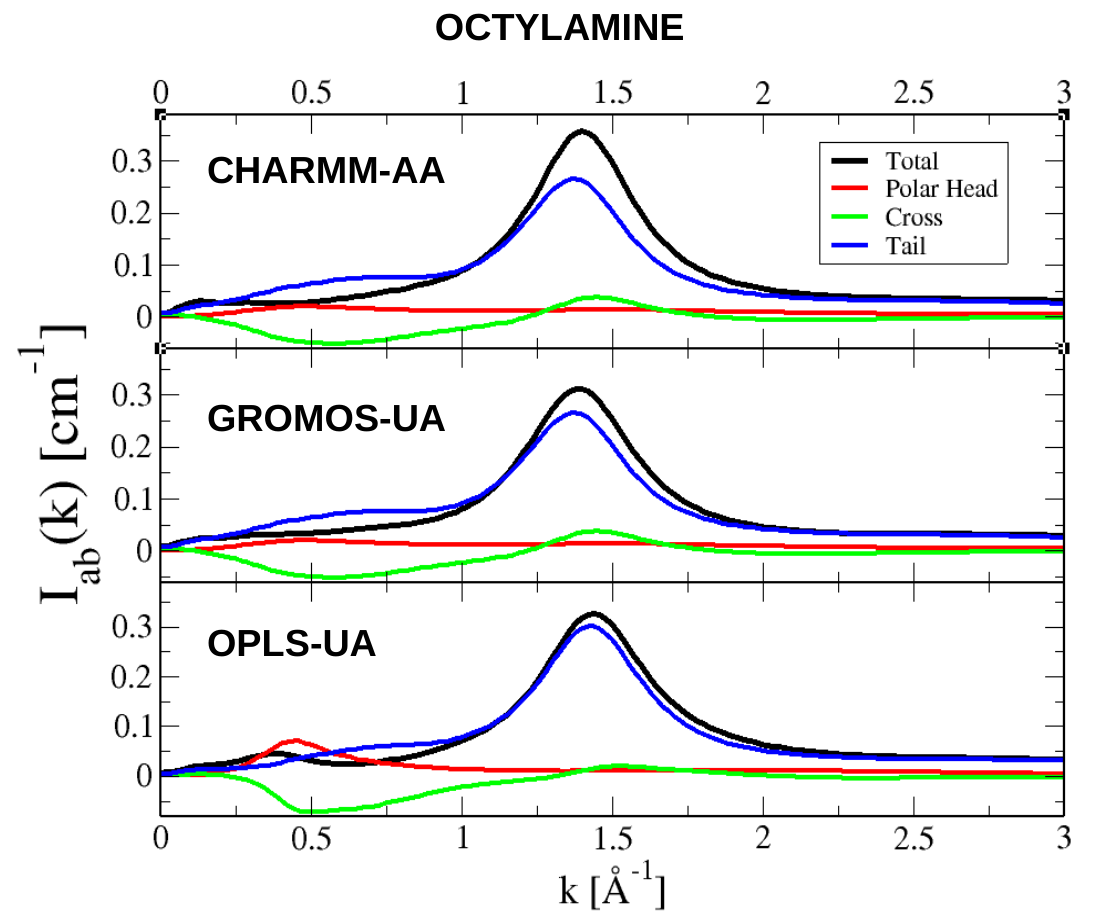}

\caption{Partial scattering contributions for octylamine}

\label{FigPartialScatt8} 
\end{figure}

A similar picture is shown for octylamine in Fig.\ref{FigPartialScatt8}.
While the total intensity remains nearly the same as for propylamine,
we notice that the partial contributions are quite different. In other
words, the quasi similarities between all propylamine scattering patterns
noticed from Fig.\ref{Fig-Ik-comp} hides large differences between
the contributions from the different molecular parts of different
amines.

Going into details, we see that it is the alkyl tail which contributes
mostly to the main peak, with, however, significant contributions
from both the polar and cross parts. These contributions differ from
one model to another, and appear more important for the all-atom models.
Perhaps the most interesting part is canceling contribution from the
cross and polar/non-polar contributions to the pre-peak, which are
more important for the OPLS model. These cancellations are typical
of charge ordering, as will be shown from the study of the pair correlation
and structure factors in the next sub-section. This type of cancellation
occurs in ionic liquids\cite{PERERA2015243} and more particularly
in room temperature ionic liquids which have been the focus of recent
studies from charge ordering perspective\cite{Ionic_1,Ionic_2,Ionic_3}.

\subsection{Study of the order parameters}

Herein, the site-site correlation functions and corresponding structure
factors are considered as generalised Landau-type order parameters
for complex disorder liquids. The order they witness is not a global
order, as in true phase transition, but the local order generated
by charge ordering, which differs from that of simple disorder liquids\cite{AUP_Charge_ordering_prepeak_neat_alc,AupPAC}
. In charge ordering, the alternation of the positive and negative
charges leads to a typical dephasing between the $+-$ correlations
and the $++\--$ correlations \cite{AupDomainOrdering,PERERA2015243}.
When charges are tied into molecules, charge order will condition
the micro-structure at molecular level. It is an open question whether
or not the resulting site-site correlations will still obey this dephasing,
and to what extent. In mono-ols, it was shown that charge order and
dephasing of charge-charge correlations are still respected. Since,
from the studies above, it appears that the amine groups tend to show
less charge order than mono-ols and favour mostly dimers, it is interesting
to examine in Fig.(\ref{FigGr_NN}) the charge order of the amine
head group nitrogen N and hydrogen atoms H, charged negatively and
positively, respectively.

\begin{figure}[H]
\includegraphics[width=0.9\textwidth]{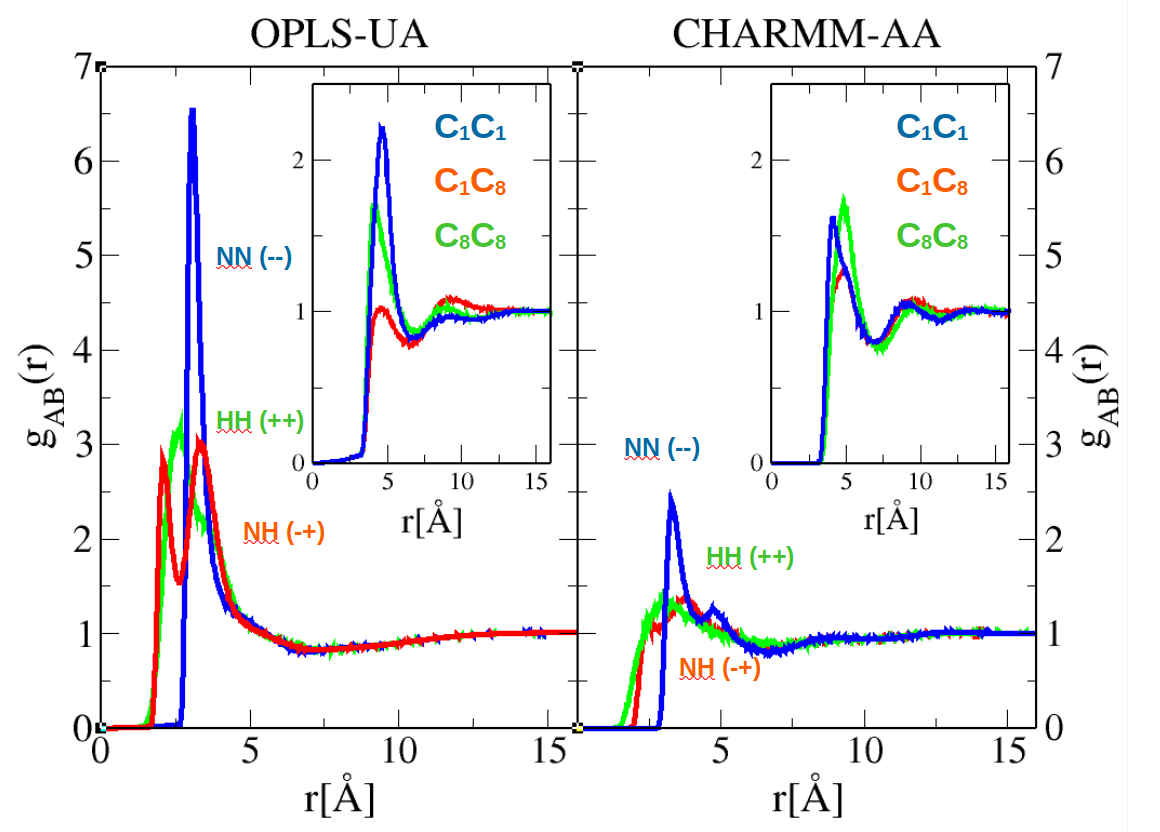}

\caption{Charge order of the amine $NH_{2}$ head group atoms for octylamine,
from the point of view of the site-site correlation functions. The
inset shows the correlations combinations between the first $C_{1}$
and last carbon $C_{8}$ groups. See text for more details.}

\label{FigCO-GR-OCTYL} 
\end{figure}

Since the OPLS model (left panel) shows a marked pre-peak similar
to that in mono-ols, we observe a clear charge ordering, with the
typical phase opposition between the N(-)H(+) correlation and the
H(+)H(+) or N(-)N(-) correlations. Incidently, because of the double
hydrogens, $g_{NH}(r)$ has a double peak since the amine head can
equally bond with the 2 hydrogen atoms. This is because the alkyl
tail blocks the tetrahedral order which would allow to separate the
bonding with the 2 hydrogens. This is a topological constraint. The
CHARMM model (right panel) has very clearly less marked charge order
correlation, seen through the lesser amplitude of the peaks, altough
these are there and obey similar dephasing. The $g_{NN}(r)$ has a
main peak with a smaller shoulder peak around $r\approx5\mathring{A}$,
which is probably a second hydrogen bonding with another nitrogen.
This is also a consequence of the topological constraints, modified
by the less localized bonding possibility, precisely because of the
weaker bonding than in OPLS.

Both insets in Fig.(\ref{FigCO-GR-OCTYL}) represent the carbon-carbon
correlations between the first carbon $C_{1}$ and the last one $C_{8}$.
Interestingly, the corresponding correlations are very Lennard-Jones
(LJ) type, with all the peaks in phase. We note a very important feature:
the first neighbour correlations are depleted, since all the curves
are slightly below the asymptote 1: this is very different than for
LJ mixtures. This second feature is very similar to that seen in the
$g_{OO}(r)$ of alcohols\cite{Pozar_2024}. It indicates that the
first and last carbon atoms are anti-correlated, a feature also seen
in neat alcohols \cite{2020_Linear_alcohols}\cite{Pozar_2024}. We
also note another important feature in the right panel inset, that
the correlation peaks for the CHARMM model between different atom
pairs are slightly out of phase, although not as much as those in
the main panel. This is a consequence of that the carbon and hydrogen
atoms in the CHARMM model bear tiny charges

\begin{figure}[H]
\includegraphics[width=0.9\textwidth]{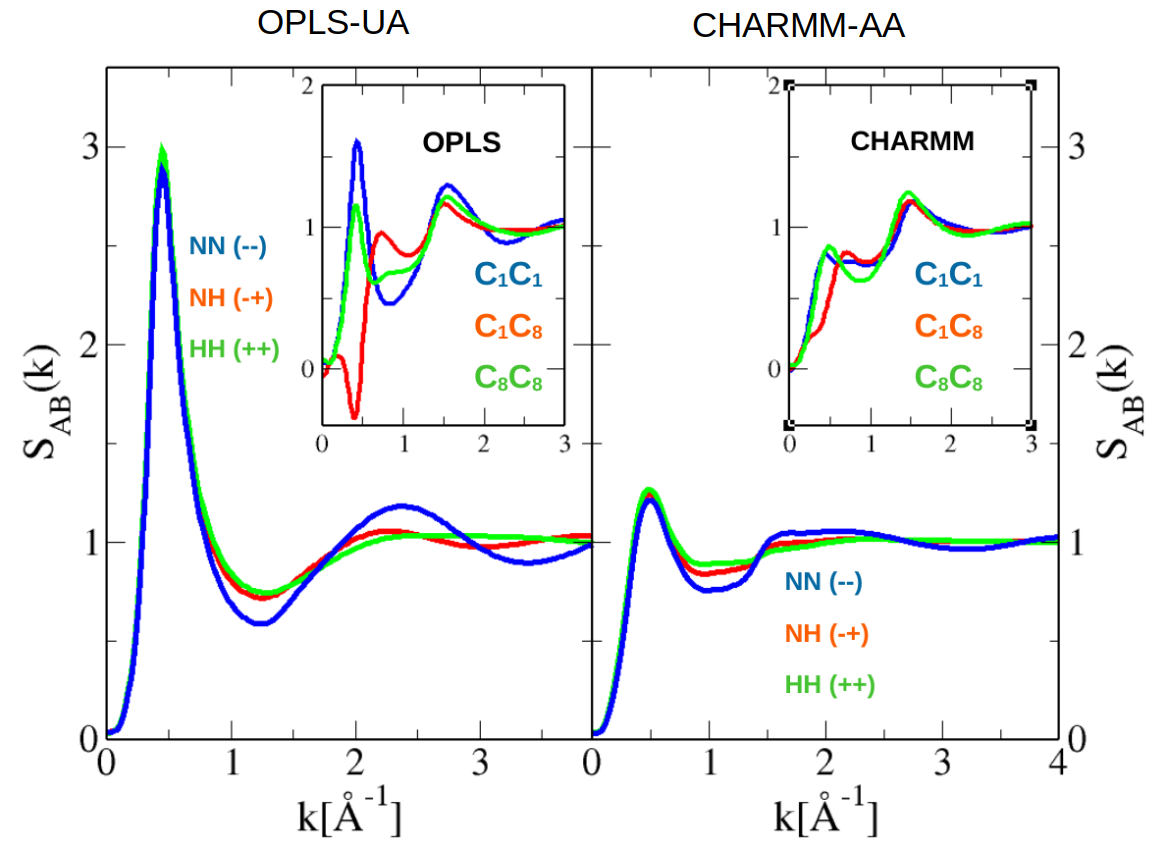}

\caption{Charge order of the amine head group atoms for pentylamine, from the
point of view of the site-site structure factors. The inset shows
the structure factors for the first $C_{1}$ and last $C_{8}$ carbon
groups. See text for more details. }

\label{FigCO-SK-OCTYL} 
\end{figure}

Fig.\ref{FigCO-SK-OCTYL} shows the atom-atom structure factors corresponding
to the correlations in Fig.\ref{FigCO-GR-OCTYL}, for both models.
Again, the similarity with the same correlations in 1-octanol are
really striking (see Fig.SI-1 in the SI document). We note that all
amine group atoms pairs have the same pre-peak, which is more marked
for OPLS than for CHARMM, as expected. This is a consequence of the
fact that all atoms in the amine group pilot the charge ordering in
the same way. Even this lay lead to small detail in the $r$-space
correlations between near neighbours, the collective effect\textcolor{blue}{{}
}is very similar. This feature justifies the need to look at both
the $r$-space and reciprocal space correlation order parameters.

It is also important to remind the physical and mathematical origin
of the pronounced pre-peak, which we have introduced in Ref.\cite{AUP_Charge_ordering_prepeak_neat_alc},
and observed in subsequent works\cite{2020_Linear_alcohols,2021_Octanols}.
In these systems we are concerned with, scattering pre-peaks are generated
by chain clustering of the hydrogen bonding groups (OH in alcohols
and NH in amines). This particular form of clustering can be seen
from 2 features in the pair correlation functions. The first feature
is a narrow and prominent first peak, which has two meaning: the height
indicates the strength of the H-bonding, while the width indicates
the ``looseness'' of the H-bonding directionality. The second feature
is the relatively shallow depletion following this first peak, and
which extends far beyond it. These 2 features can be clearly seen
in the left panel of Fig.\ref{FigCO-GR-OCTYL} for the OPLS $g_{NN}(r)$,
and they are much less marked for the CHARMM model. From this, we
are able to conclude that the OPLS model has more and longer N-H chain
clustering than the CHARMM model, which might explain the more prominent
scattering pre-peak.

We also note that the depletion correlations of the carbon atoms produce
pre-peaks and anti-peak in the case of the OPLS model, again indicating
the strong correlation between carbon atoms, which is a direct consequence
of the fact that these are tied to the amine groups.

\begin{figure}[H]
\includegraphics[width=0.9\textwidth]{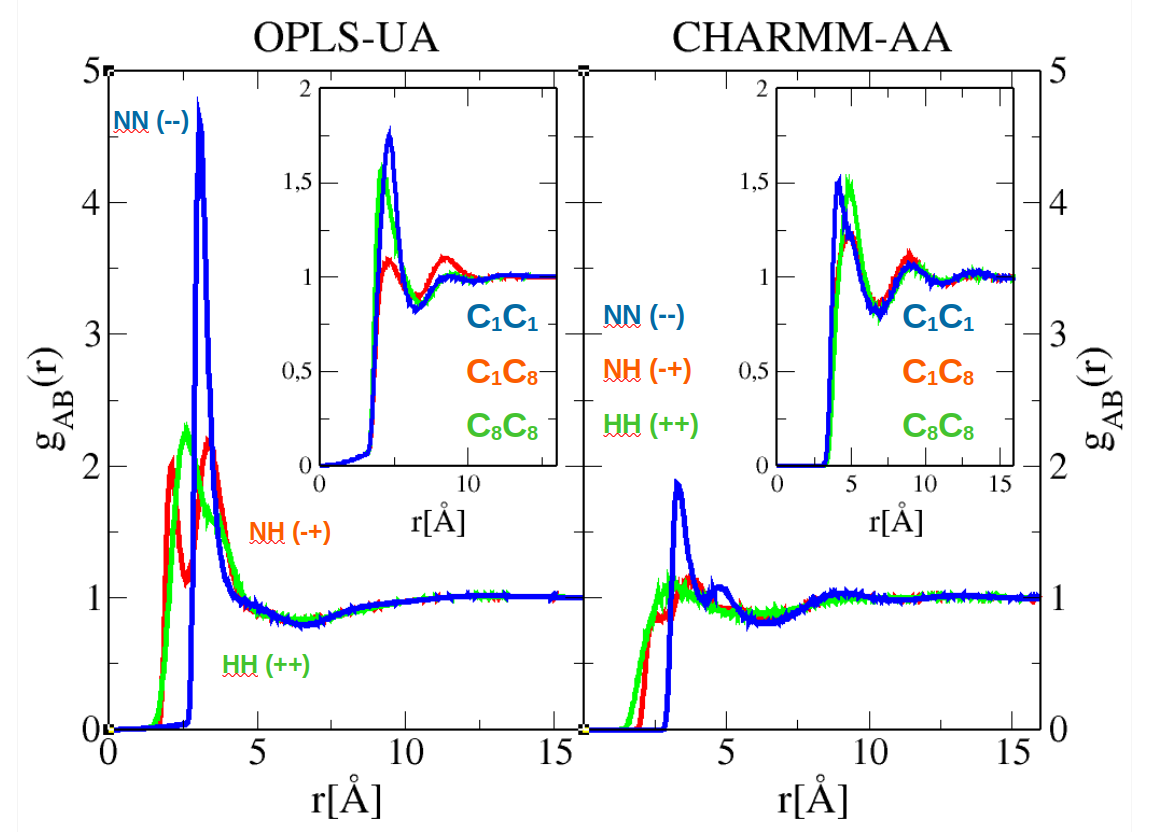}

\caption{Charge order of the amine head group atoms for pentylamine, from the
point of view of the site-site correlation functions. The inset shows
the correlations combinations between the first $C_{1}$ and last
carbon $C_{5}$ groups. See text for more details.}

\label{FigCO-GR-PENT} 
\end{figure}

\begin{figure}[H]
\includegraphics[width=0.9\textwidth]{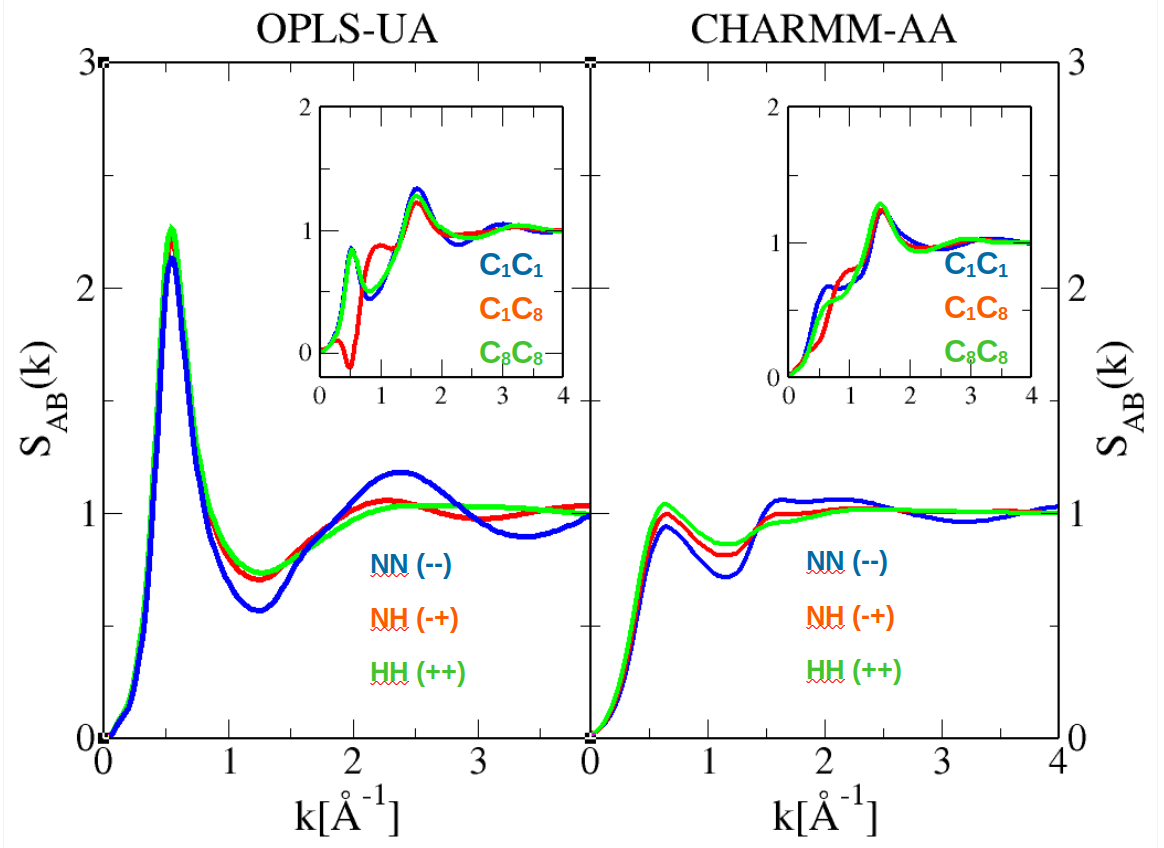}

\caption{Charge order of the amine head group atoms for pentylamine, from the
point of view of the site-site structure factors. The inset shows
the structure factors for the first $C_{1}$ and last $C_{5}$ carbon
groups. See text for more details. }

\label{FigCO-SK-PENT} 
\end{figure}

These features are shared by all amines, with variations in amplitude
of the effects. For instance, in Fig.\ref{FigCO-GR-PENT} and Fig.\ref{FigCO-SK-PENT},
we illustrate this similarity in the case of correlations in pentylamine.

It can be seen that these figures are very similar to the previous
ones, except for a general lesser amplitude of the charge ordering.
This similarity indicates two features, that we previously noticed
in mono-ols. Firstly, the amine group correlations are strikingly
similar, with only small amplitude difference, despite the underlying
differences between models. Secondly, the connections between the
hydrogen bonding head groups and the quasi neutral tails is very important,
despite the fact that the tails are expected to be more or less decorrelated
around the polar head cluster.

In connection with the partial scattering contributions of Fig.\ref{FigPartialScatt8},
one may ask why, if charge order between the alkanols and the alkylamines
look quite similar, there is no pre-peak in the amines? It appears
that the amine group correlations are not so strong, while the alkyl
correlations, including their anti-correlations, are as strong as
in alkanols, hence over canceling the positive correlations that would
give pre-peak. This conclusion is entirely supported by the OPLS model,
which precisely has amine group correlations very similar to that
of the hydroxyl group of the alkanols.

\section{Force field dependence of the micro-structure of alkylamines}

One important difference between the 3 force field models is the all-atom
modeling of the alkyl tail of the CHARMM model, with partial charges,
albeit quite small, on the carbon and hydrogen atoms. However, the
GROMOS model has also partial charges in the alkyl tail. This is the
main difference with the OPLS model, which has zero charges on the
$CH_{n}$ groups. As seen in the previous section, the absence of
charges on the alkyl tail does not necessary imply that the alkyl
correlations are weak (see the insets of Figs.\ref{FigCO-GR-OCTYL}
to Fig.\ref{FigCO-SK-PENT}). As stated previously, it is in fact
the balance of correlations and anti-correlations between the head
and tail groups which determines the existence and the strength of
the pre-peak. Apparently, the higher valences of the OPLS model somehow
compensate the absence of charges on the tail atoms.

\begin{figure}[H]
\includegraphics[width=0.9\textwidth]{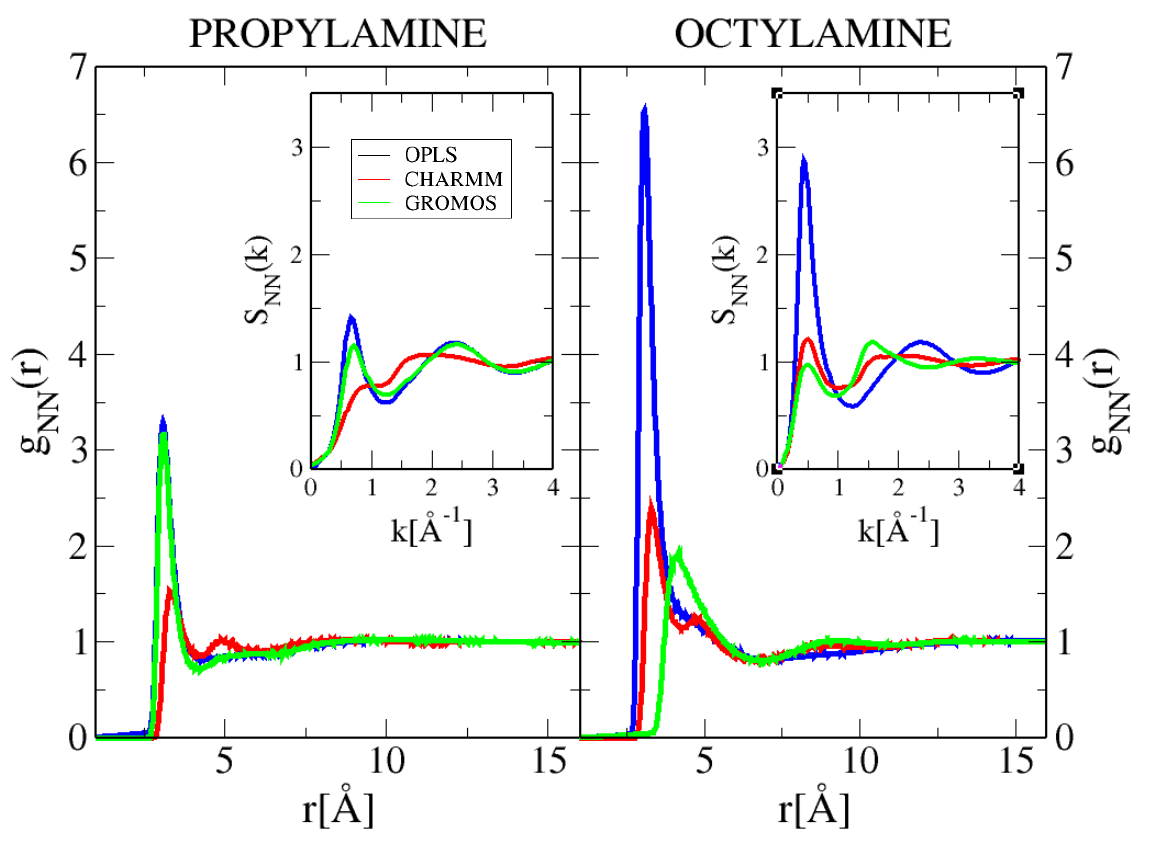}

\caption{Nitrogen-nitrogen correlation function for propylamine (a) and octylamine
(b). The corresponding structure factors are shown in the insets.}

\label{FigGr_NN} 
\end{figure}

We compare below the correlations of the nitrogen atom, which is central
to the hydrogen bonding, between the 3 models. We focus on the amines
with the smallest tail, which is propylamine, and the longest one,
octylamine. Fig.\ref{FigGr_NN} shows the $g_{NN}(r)$ for all 3 models,
for propylamine and octylamine as well as the corresponding structure
factors in the insets.

We note that there are important differences between the short range
correlations of the alkylamines, which are common to all 3 models:
there is a widening of the base of the first peak for the longer alkylamine,
indicating that N-N contacts are both more enhanced for direct dimer
pair formation (higher first peaks) and allowing higher n-mer formation,
such as most probably trimers. This interpretation is supported by
the larger correlation depletion range at higher distances, indicating
that the separation with next N neighbours increases with alkyl tails.
This is very counter-intuitive, since one would expect larger N clusters
when the alkyl tails are entropically favorable, that is when they
are small. Instead, it would appear that longer tails favour micelle-type
at the core. In other words, longer alkyamines would tend to behave
like micellar melts.

We note that this micellisation-like behaviour is less supported by
the GROMOS model, since the main peak of $g_{NN}(r)$ is smaller,
but also the pre-peak of $S_{NN}(k)$ is smaller than the other models.

It is quite striking that the large disparity between models at the
level of the site-site correlations, do not reflect so much on the
total $I(k)$. This is because of the cancellation mechanisms explained
in Section 4.2. It supports the idea that there are many local molecular
arrangement possible that lead to similar observables.

\section{Charge ordering differences between water, amines and alcohols}

It is instructive to examine the charge ordering mechanism by comparing
three different types of hydrogen bonding liquids, water, amines and
alcohols. We note that the water molecule OH2 has a C2v symmetry in
the Sch\"{o}nflies notation\cite{morawiec2022indexing}, which is also
that of the NH2 amine head group. This symmetry favours branching
patterns. In contrast, the OH hydroxyl head group has linear symmetry,
which favours chaining patterns. Both symmetries are lost when attached
to the alkyl tails. However, the bonding patterns inherit these respective
symmetries. This is the reason why the cluster probabilities in Fig.\ref{Pn}
look very much like that for water, and do not have the typical peak
at cluster size $s=5$ observed for all alcohols \cite{2020_Linear_alcohols,2021_Octanols}.
The amines don't maintain the branching characteristics of C2v because
of the tail. However, the C2v symmetry definitely hinders chaining
patterns and makes small clusters (dimers and trimers) likely.

\begin{figure}[H]
\includegraphics[scale=0.3]{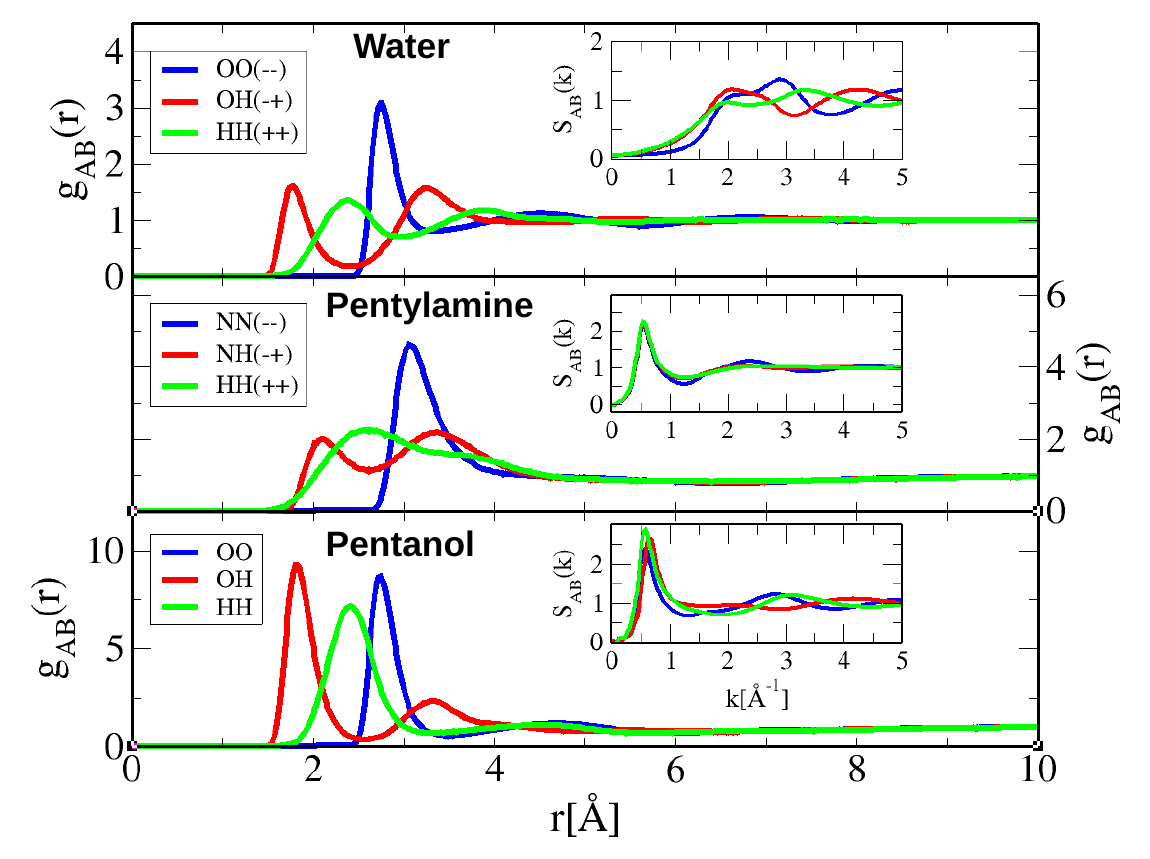}

\caption{Compare the charge order parameters between SPC/E water (top panel),
OPLS pentylamine (middle panel) and pentanol (lower panel). The respective
structure factors are shown in the insets. Note that the vertical
scales differ for each panels and insets.}

\label{Fig-COMP-WAA} 
\end{figure}

Fig.\ref{Fig-COMP-WAA} compares the charge order parameters between
water (SPC/E model) and pentylamine (OPLS model), base on the identity
of the Sch\"{o}nflies C2v symmetry of the OH2 and NH2 groups. The comparison
illustrates several appealing symmetries. For comparison, the same
order parameters for pentanol (OPLS) are equally shown in the lower
part.

The similarities between the pair correlation order parameters between
water and the amine are really striking in the short range parts in
the $r$-range $[0-5\mathring{A}]$ where the first peaks are most
apparent. The NN and NH shapes are quite similar to the OO and OH
shapes. It is only the HH part which differs significantly, indicating
that the alkyl tail perturbs the positioning of the hydrogen atoms,
as opposed to the case of water where they are not hindered. This
is very different from the pentanol, where the OH bonding has a clear
strength in magnitude, favoured by the direct OHO chain bonding mechanism
observed in all acohols\cite{2020_Linear_alcohols}, including those
with branching tails\cite{2021_Octanols}. These findings supports
and confirms the symmetry argument provided above.

The depletion range part, in the $r$-range $[5\mathring{A}-10\mathring{A}]$
contains the depletion correlations, which do not exist for water,
and which are not much visible in Fig.\ref{Fig-COMP-WAA} for the
amine and the alcohol, because of the squashed vertical scale, but
are visible for instance in Fig.\ref{FigGr_NN} above, and also in
our paper in alcohols microstructure\cite{Pozar_2024}, and in Fig.SI-X
of the SI document. This depletion range is more marked for the alcohol
than for the amine (see Fig.SI-XX). This is compatible with the prominent
scattering pre-peak in alcohols, as opposed to weak ones in amines.

However, the structure factor order parameters totally fail this similarity
argument, which seems to hold only in real space -- where actual
bonding occurs and spatial topologies are established. In the reciprocal
space, it is more the global structure that is apparent, and clearly
water and amines are not the same type of liquid. This explains the
strong differences in reciprocal space. This is apparent from the
near in phase correlations for the amine, and the out-of-phase correlations
for water. In that, the alcohol appears as closer to the amine, and
water really stands apart.

\section{Discussion}

This study demonstrates that presence or absence of scattering pre-peak
is decoupled from the presence or absence of clusters in the system.
Similarly, a direct cluster study may or may not reveal specific clusters.
It is really the order parameters, which are the correlation functions,
which reveal or not if there is clustering, depending if some pre-peak
and anti-peaks in the atom-atom structure factors cancel or not. These
functions are unfortunately not available from experiments (except
perhaps in some very specific small angle studied of neutron scattering\cite{Soper_epsr,EXP_SANS_Soper_pure_methanol}).
In other words, it is necessary to perform computer simulations, which
means introducing approximate force fields, and some arbitrariness.

We come back to the remark in Section 3.1 about the possible reasons
for the better agreement of the scattering intensities with simulation
data for alcohols\cite{2020_Linear_alcohols} instead of amines. We
would like to point out that, in fact, the differences between the
experimental and simulations $I(k)$ are always there, and can be
more or less prominent, depending on models. But, the fact that alcohols
show a better trend for pre-peak help mask these differences. This
is in relation with the cancellation feature due to the underlying
charge ordering, that is always at the origin of the pre-peak. We
conjecture that systems which show less or no pre-peak, are more prone
to concentration fluctuations, a feature which tends to smooth out
aggregates, as opposed to those who has marked pre-peaks, which have
more robust and long lived aggregate structures. This can be seen
as a classical equivalent of the boson-fermion symmetry, something
that we have already observed in a previous recent work\cite{my_aqdiox}.

This type of cancellation between different atom-atom structure factor,
and driven by the charge order mechanism, may play an important role
in the x-ray study of large molecules in soft-matter systems and those
of biological interest, where even larger molecules such as enzymes
for instance, are mixed with solvent and other smaller molecules\cite{multiple_prePeaks}.
In such cases, the disparity of sizes induce concentration fluctuations,
which tend to raise the small-$k$ part of the scattering patterns.
Then, in the absence of the pre-peak, it is difficult to decide if
the large scattering amplitudes at small $k$ are a signature of concentration
fluctuations alone, or hide significant clustering hidden to the experimental
observation. This is for instance the case in aqueous t-butanol mixtures\cite{meso-inhom1},\cite{meso_aggreg2},
or aqueous mixtures of small surfactant molecules\cite{Darrigo-Teixeira,EXP_SANS_Texeira_Water_DIOLS,EXP_SANS_Texeira_Water_DIOLS_TEMP}.
Our present results incite to more precautions in interpreting these
data, and may require revisiting many of the previous scattering results
in soft-matter systems.

\section{Conclusion}

In this work, we have proposed to consider charge order as a new form
of local order, associated to the atom-atom pair correlation functions
and corresponding structure factor. We have illustrated the usefulness
of this approach to understand the apparent weakness of clustering
in liquid amines, as suggested by the very weak x-ray scattering pre-peaks.
The computer simulation studies of several force fields have revealed
that significant H-bonding and clustering occurs in liquid amines,
but that charge ordering associated to the particular symmetries of
the associating atomic groups, tend to produce canceling contributions
which diminish the height of the scattering intensities of the pre-peak.
However, significant differences between different force fields are
also found, which tend to affect the interpretation. In the case of
liquid amines, no particular force field appears are reliable, although
the overall agreement is not so bad.  Moreover, all force field converge
as far as interpretations of the microscopic process are concerned,
which is the positive side of the simulation approach.

\section*{Acknowledgements}

We thank DELTA for providing synchrotron radiation at beamline BL2
and technical support. The help of Jaqueline Savelkous, Eric Schneider,
and Dirk L\"{a}tzenkirchen-Hect is thankfully aknowledged for preparatory
measurements at BL8. This work was supported by the BMBF via DAAD
(PROCOPE 2024-2025, Project-ID 57704875) within the French-German
collaborations PROCOPE (50951YA), \emph{Analysis of the molecular
coherence in the self-assembly process: experiment and theory}.

 \bibliographystyle{jpcb_final.bst}
\bibliography{neatamines}

\end{document}